\documentclass[conference]{IEEEtran}
\IEEEoverridecommandlockouts
\usepackage{cite}
\usepackage{amsmath,amssymb,amsfonts}
\usepackage{graphicx}
\usepackage{textcomp}
\usepackage{fourier}
\usepackage{graphicx}
\usepackage{url}
\usepackage{wrapfig}
\usepackage{breakcites}
\usepackage[T1]{fontenc}
\usepackage{times}
\usepackage{array, makecell} %
\usepackage{algorithm2e}
\usepackage{algpseudocode}
\usepackage[dvipsnames]{xcolor}
\usepackage{nicematrix}
\def\BibTeX{{\rm B\kern-.05em{\sc i\kern-.025em b}\kern-.08em
    T\kern-.1667em\lower.7ex\hbox{E}\kern-.125emX}}
\begin{document}

\title{AudRandAug: Random Image Augmentations for Audio Classification}

\author{\IEEEauthorblockN{1\textsuperscript{st} Teerath Kumar}
\IEEEauthorblockA{\textit{CRT-AI Centre,
School of Computing} \\
\textit{
Dublin City University}\\
Dublin, Ireland  \\ teerath.menghwar2@mail.dcu.ie}\\
\and
\IEEEauthorblockN{2\textsuperscript{nd} Muhammad Turab}
\IEEEauthorblockA{\textit{} \\
\textit{
Norwegian University of Science and Technology (NTNU)}\\
Gjovik, Norway \\
turabbajeer202@gmail.com }
\and
\IEEEauthorblockN{3\textsuperscript{rd}Alessandra Mileo}
\IEEEauthorblockA{\textit{INSIGHT Research Centre,
School of Computing} \\
\textit{
Dublin City University}\\
Dublin Ireland }
\and
\IEEEauthorblockN{4\textsuperscript{th} Malika Bendechache}
\IEEEauthorblockA{\textit{Lero Research Centre,
School of Computer Science} \\
\textit{
University of Galway}\\
Galway,  Ireland }
\and
\IEEEauthorblockN{5\textsuperscript{th} Takfarinas Saber}
\IEEEauthorblockA{\textit{Lero Research Centre,
School of Computer Science} \\
\textit{
University of Galway}\\
Galway,  Ireland }

}

\maketitle

\begin{abstract}
Data augmentation has proven to be effective in training neural networks. Recently, a method called RandAug was proposed, randomly selecting data augmentation techniques from a predefined search space. RandAug has demonstrated significant performance improvements for image-related tasks while imposing minimal computational overhead. However, no prior research has explored the application of RandAug specifically for audio data augmentation, which converts audio into an image-like pattern. To address this gap, we introduce AudRandAug, an adaptation of RandAug for audio data. AudRandAug selects data augmentation policies from a dedicated audio search space. To evaluate the effectiveness of AudRandAug, we conducted experiments using various models and datasets. Our findings indicate that AudRandAug outperforms other existing data augmentation methods regarding accuracy performance. 
\end{abstract}


\section{Introduction}

Deep learning (DL) has successfully addressed complex problems, proving proficiency in managing large datasets and discerning intricate patterns. Consequently, DL has become an indispensable tool for various tasks, including image processing \cite{aleem2022random,wu2015image,fujiyoshi2019deep,kumar2021class,kumar2023advanced,kumar2023rsmda,ranjbarzadeh2023me,roy2023wildect,khan2022introducing, chandio2022precise, singh2023deep,singh2023understanding}, natural language processing \cite{torfi2020natural,hirschberg2015advances}, and audio processing \cite{hershey2017cnn,fu2010survey,turab2022investigating,park2020search,chandio2021audd,sarwar2022advanced} and other DL application~\cite{roy2022computer, kumar2022stride,  khan2022data, irfan2023go, steinberg2010method,mandal2016static, mandal2023measuring, mandal2023multimodal, mandal2016static}. Notably, DL has demonstrated impressive performance in the field of audio data analysis. Extensive research has been conducted on numerous tasks such as audio classification, music generation, and environmental sound classification \cite{lee2017sample}.

Previous studies \cite{fu2010survey,turab2022investigating,park2020search,kumar2020intra} have highlighted the challenge of training neural networks directly on raw audio data, as it can be difficult for them to learn essential features. To overcome this limitation, researchers have shown that neural networks can achieve significantly improved performance by training them on audio-specific features \cite{palanisamy2020rethinking}. Convolutional Neural Networks (CNNs) have been widely employed for audio content analysis, utilizing various features and methods \cite{turab2022investigating,palanisamy2020rethinking,li2019multi}.

\begin{figure}
\centering 
\includegraphics[width=0.5\textwidth, height=6cm]{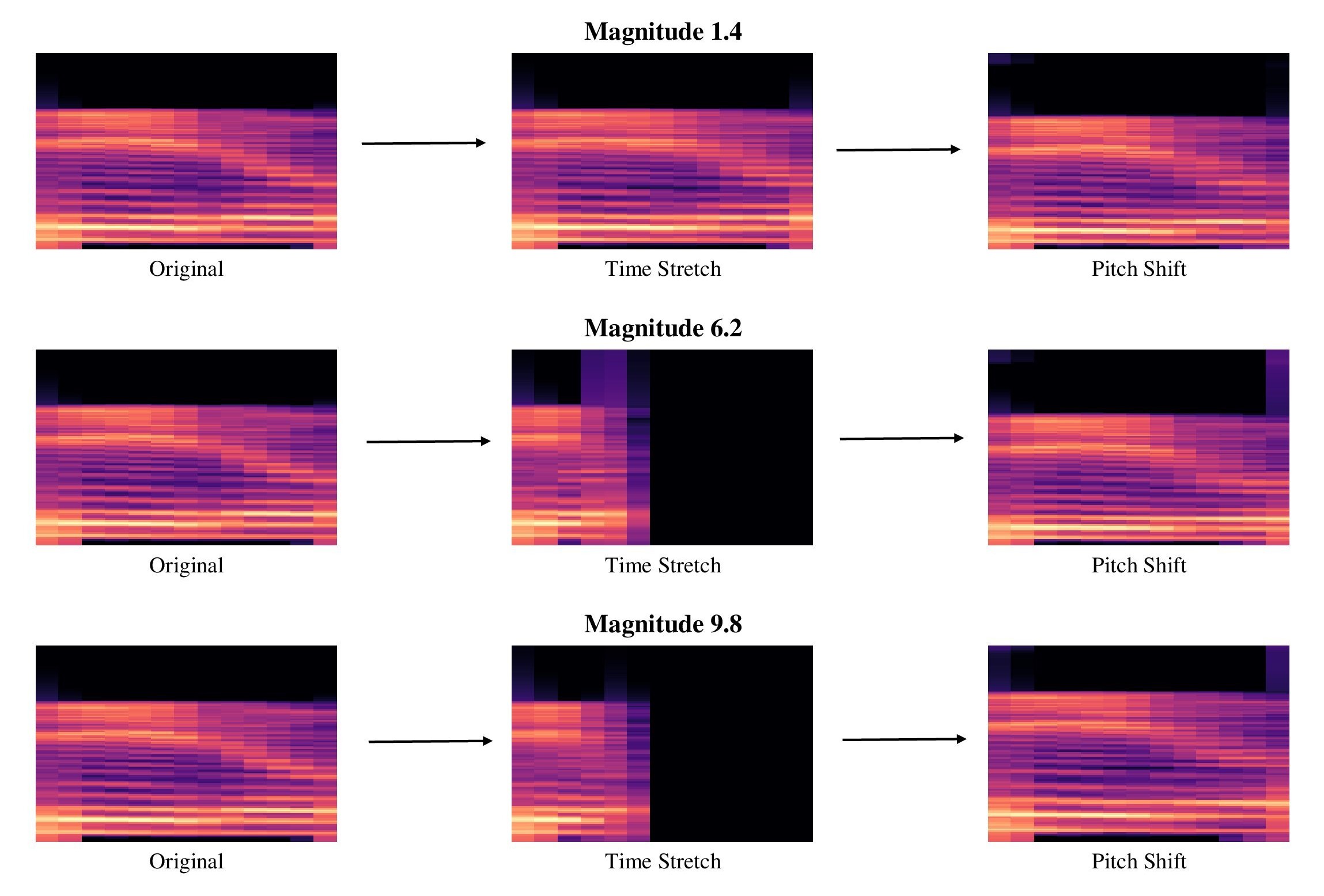}
\caption{\textbf{Example audio mel-spectrograms augmented by AudRandAug.} In these examples, N = 2 and different M magnitude values are shown. As the magnitude of the distortion rises, so does the strength of the augmentation.} \label{fig:audrandaug_images}
\end{figure}

Despite the accuracy achieved through feature extraction methods, there remains room for improvement due to limited availability of labeled data. Deep learning models require large-scale labeled data to learn more accurate features. However, the process of labeling data on a large scale is tedious, time-consuming, and expensive \cite{kumar2021binary}. To address this challenge, various data augmentation (DA) techniques can be applied to existing data by increasing the diversity and size of data, allowing the model to learn from different perspectives of each sample. The objective is to train the network on additional distorted data, enabling the network to become invariant to these distortions and generalize better to unseen data. Several studies have explored data augmentation methods in the audio domain \cite{ko2015audio,nanni2020data,jain2021spliceout}. In line with the principles of image, randAug \cite{cubuk2020randaugment}, we propose a novel approach for audio classification called AudRandAug, which is demonstrated in Figure~\ref{fig:audrandaug_images}. Our work contributes in the following ways:

\begin{itemize}

\item Inspired by RandAug, we introduce a novel data augmentation technique named AudRandAug.
\item We perform several experiments to select the most effective augmentation methods for inclusion in the search space of AudRandAug
\item To validate the proposed approach, we perform several experiments on different datasets.
\item We provide code in GitHub repository: \url{https://github.com/turab45/AudRandAug.git}
\end{itemize}


The rest of the paper is organized as, section~\ref{related_work} discusses the related work, section~\ref{methodology} explains the proposed methodology, section~\ref{experimental_design} provides experimental details such as datasets, training setup and results, and finally section~\ref{conclusion} concludes  the work. 
\section{Related Work}\label{related_work}

This section discusses relevant data augmentation work in the audio domain. Deep learning methods have been widely applied to audio/sound data, such as music genre classification \cite{dong2018convolutional,choi2017convolutional,zhang2016improved}, audio generation  \cite{oord2016wavenet,roberts2018hierarchical}, environmental sound classification \cite{guzhov2021esresnet,aytar2016soundnet,demir2020new}, and more \cite{chachada2014environmental,dandashi2017survey}. From an architectural perspective, various methods have been explored for audio classification. Models using 1-D Convolution, such as EnvNet~\cite{tokozume2017learning} and Sample-CNN \cite{lee2017sample}, have been proposed for raw audio waveform classification. However, recent work has primarily focused on utilizing CNN on spectrogram (an image pattern), which has led to state-of-the-art (SOTA) results. Dong et al. \cite{dong2018convolutional} proposed a CNN-based method for music genre classification, and Palanisamy et al. \cite{palanisamy2020rethinking} showed that an ImageNet pre-trained model could be a strong baseline network for audio classification.

In addition to architectural considerations, data augmentation has shown promising results in various audio tasks. For convenience, audio data augmentation can be broadly divided into two levels: (i) data augmentation on the raw audio level and (ii) data augmentation on the feature level.

\begin{table*}[hpt!]
    \centering\renewcommand\cellalign{lc}
\setcellgapes{3pt}\makegapedcells

    \begin{tabular}{|l|l|}
    \hline 
         \textbf{Augmentation} &  \textbf{Description} \\ \hline 
         Noise Injection &  This involves introducing additive white Gaussian noise  (AWGN) to the   original audio \\  & recording through element-wise addition. \\ \hline
         Pitch Shifting &  This alters the pitch of an audio recording without  impacting its duration  or timing \\ \hline
         Time Stretching &  
This modifies the speed or duration of an audio recording  while  preserving   its pitch and tonal  \\ & characteristics. This is achieved by utilizing the  Short-time Fourier   transform (STFT) technique. \\ \hline
         Padding &  Padding in audio refers to the technique of enhancing the sound quality of a recording \\ & by replacing   a fraction of the beginning or end of the audio with padded sections.  \\ \hline
         Clip &  Clipping removes excessive audio signal to prevent distortion and ensure a clean sound.\\  \hline
         Reverse & Reversing an audio signal involves inverting its polarity, commonly used to create a reversed  \\ &  playback effect or special audio effects.\\ \hline
         Band Pass Filter &  
A band pass filter is an electronic filter designed to permit a specific range of frequencies to \\ &  pass through while attenuating all others. This filter is frequently employed in audio applications \\ &  to eliminate unwanted noise and interference, ensuring optimal sound quality. \\ \hline
         Gain &  To enhance the model's resilience to variations in input gain, it is beneficial to multiply the  \\ &  audio by a random amplitude factor. By doing so, the model becomes less dependent on specific  \\ &   gain values and exhibits more consistent performance across a diverse  range of input signals. \\ \hline
         Time Masking & This is an audio technique where a randomly selected portion of the audio is made silent,  \\ &  effectively removing unwanted noises or creating unique effects. This method is  commonly  \\ & employed to enhance audio quality and achieve specific audio effects \\ \hline
         
    \end{tabular}
    \caption{All the used data augmentation methods }
    \label{tab:augmentation}
\end{table*}

\subsection{Data augmentation on raw audio level}
Extensive research has been carried out on using deep learning techniques for raw audio data analysis. Various models have been developed specifically for classifying raw audio waveforms using 1-D Convolutions. For instance, EnvNet~\cite{tokozume2017learning} and Sample-CNN~\cite{lee2017sample} are notable examples of models that leverage raw audio waveforms as their inputs. These models have demonstrated significant advancements in achieving SOTA performance across different sound categories~\cite{lee2017raw}.

\subsection{Data augmentation on features level}
Recent research has emphasized employing CNNs on spectrograms to achieve SOTA outcomes. Dong et al.~\cite{dong2018convolutional} proposed a CNN-based method for music genre classification, achieving accuracy of 70\%. Additionally, Palanisamy et al.~\cite{palanisamy2020rethinking} demonstrated that a pre-trained ImageNet model can serve as a strong baseline network for audio classification. To further enhance generalization, a few studies have explored feature extraction~\cite{turab2022investigating,su2019environment,liu1998audio} and data augmentation approaches ~\cite{ko2015audio,nanni2020data,kim2021specmix}. In the work by Turab et al., ~ \cite{turab2022investigating}, different feature selection methods were investigated for audio using ensemble techniques. The search for optimal augmentation policies was explored in~ \cite{BSGHC3_2020_v25n6_854}, while Kumar et al.~\cite{kumar2020intra} proposed a novel intra-class random erasing data augmentation to enhance network robustness. Furthermore, Kim et al.~\cite{kim2021specmix} introduced Specmix, a novel audio data augmentation technique specifically designed for time-frequency domain features. This approach improved the performance of various neural network architectures by up to 2.7\%. Salamon et al. ~\cite{salamon2017deep} proposed a deep neural network architecture coupled with audio augmentations to address the challenge of data scarcity in their work. Among all these approach, none has explored image randAugment approach for audio data. To the best of our knowledge, we are the first to explore it. 

\section{Proposed Methodology}\label{methodology}
In this section, we explain the proposed method and used data augmentation methods in the search space. 
\subsection{Method}
Inspired by the RandAug~\cite{cubuk2020randaugment} in the domain of images, we introduce a random data augmentation technique for audio classification called AudRandAug. This approach involves determining the optimal parameters for each specific data augmentation operation. Subsequently, we apply a total of $N$ data augmentations, each with its corresponding optimal magnitude or parameter(s), like an optimal magnitude for time stretch augmentation is 1.4. The proposed algorithm is provided in \textbf{algorithm~\ref{alg:audrandaug}}.

\begin{algorithm}[hbt!]
\caption{AudRandAug}\label{alg:audrandaug}
\begin{algorithmic}[1]
\Require $N$ (Integer), $M$ (List)
\Ensure Selected augmentations and magnitudes

\State augmentations $\gets$ ["Noise", "Pitch", "Time", "Padding", "Clip", "Trim", "Reverse", "BPF", "BSF"]
\State selected\_augmentations $\gets$ random choice from augmentations, size $N$
\State selected\_magnitude $\gets$ $M$ where selected\_augmentation equals augmentation
\State \Return $[(aug, m) \text{ for } (m, aug) \text{ in } \newline \text{zip}(selected\_magnitude, selected\_augmentations)]$ where m is magnitude of particular augmentation

\end{algorithmic}
\end{algorithm}

In our proposed approach, we adopt the same algorithm as described in RandAug \cite{cubuk2020randaugment}. The selection of data augmentation from the search space is performed with a uniform probability. We investigate several essential audio data augmentations, keeping in mind that all augmentations are applied to audio waveforms and subsequently converted into Mel-spectrograms. Finally, these augmented spectrograms are used as inputs to the CNN model. Table~\ref{tab:augmentation} provides a detailed overview of each used data augmentation technique.

\section{Experiment Design}\label{experimental_design}
In this section, we explain the training setup, dataset, and results. 
\subsection{Training Set up}
We used custom CNN and pre-trained VGG model, 0.001 learning rate, Adam optimizer, 100 epoch.  The custom CNN is 2 convolutional layers network. First convolutional layer followed by max pooling, drop with 0.2. Second convolutional layer is followed by max pooling then flatten. Then two dense layers are used. 

\begin{table*}[ht] 
\centering

\begin{NiceTabular}{lcccc}[hvlines]
  \Block{2-5}{\textbf{Custom CNN Model Results}} \\
   \\ 
  \Block{2-1}{\textbf{Augmentation}} & \Block{1-2}{\textbf{FSDD dataset}} & \Block[c]{1-3}{\textbf{UrbanSound8K dataset }}   \\ 
  &  \textbf{Performance} & \textbf{Change ($\Delta D$)} & \textbf{Performance} & \textbf{Change ($\Delta D$)} \\ 
Baseline &  92.00 &  - &  95.00 &  - \\
 + Noise Injection &  94.5 &  \textcolor{ForestGreen}{2.5} &  97.27 &  \textcolor{ForestGreen}{2.27} \\
 + Pitch Shifting &  94.83 &  \textcolor{ForestGreen}{2.83}&  97.26 &  \textcolor{ForestGreen}{2.26}\\
 + Time Stretching &  92.50 &  \textcolor{ForestGreen}{0.5}&  97.23 &  \textcolor{ForestGreen}{2.23}\\
 + Padding &  92.50 &  \textcolor{ForestGreen}{0.5} &  97.13 &  \textcolor{ForestGreen}{2.13} \\
 + Clip &  93.33 &  \textcolor{ForestGreen}{1.33}&  93.11 &  \textcolor{red}{1.89}\\
 + Reverse &  93.83 &  \textcolor{ForestGreen}{1.83}&  93.13 &  \textcolor{red}{1.87}\\
 + Band Pass Filter &  93.00 &  \textcolor{ForestGreen}{1.0}&  97.31 &  \textcolor{ForestGreen}{2.31}\\
 + Gain &  96.50 &  \textcolor{ForestGreen}{4.5}&  97.32 &  \textcolor{ForestGreen}{2.32}\\
 + Time Mask &  92.16 &  \textcolor{ForestGreen}{0.16}&  96.56 &  \textcolor{ForestGreen}{1.56}\\
 +  Ours  & 97.16 & \textcolor{ForestGreen}{5.16} & 96.37 & \textcolor{ForestGreen}{1.37}  \\  


  \Block{2-5}{\textbf{VGG Model Results}} \\
   \\ 
 
Baseline &   95.95 & - &  96.37 & - \\ 
 + Noise Injection &  98.16 &  \textcolor{ForestGreen}{2.15} &  97.89 &  \textcolor{ForestGreen}{1.52} \\
 + Pitch Shifting &  98.66 &  \textcolor{ForestGreen}{2.71}&  98.19 &  \textcolor{ForestGreen}{1.82}\\
 + Time Stretching &  94.18 &  \textcolor{red}{1.77}&  91.17 &  \textcolor{red}{5.2}\\
 + Padding &  93.87 &  \textcolor{ForestGreen}{2.39} &  98.34 &  \textcolor{ForestGreen}{1.97} \\
 + Clip &  98.66 &  \textcolor{ForestGreen}{2.71}&  98.42 &  \textcolor{ForestGreen}{2.05}\\
 + Reverse &  93.83 &  \textcolor{red}{2.12} &  95.92 &  \textcolor{red}{0.45}\\
 + Band Pass Filter &  93.00 &  \textcolor{ForestGreen}{2.95}&  98.25 &  \textcolor{ForestGreen}{1.88}\\
 + Gain &  98.66 &  \textcolor{ForestGreen}{2.71}&  97.09 &  \textcolor{ForestGreen}{0.72}\\
 + Time Mask &  94.39 &  \textcolor{red}{1.56}&  97.11 &  \textcolor{ForestGreen}{0.74}\\
 +  Ours  & 98.92 & \textcolor{ForestGreen}{2.97} & 98.63 & \textcolor{ForestGreen}{2.26}  \\  
\end{NiceTabular}
    \caption{Result using custom CNN and Pre-trained VGG models}
    \label{tab:custom_cnn}
\end{table*}

\subsection{Datasets}
We use Free Spoken Digits Dataset (FSDD)~\cite{jackson2016free}  which is a simple audio dataset consisting of English spoken digit recordings in .wav files at 8khz.  It contains 3,000 recordings from 6 speakers (50 of each digit per speaker) and English pronunciations, and it has 10 classes (0-9) and duration of the recordings is 1-2 seconds. Another dataset  UrbanSound8K dataset~\cite{10.1145/2647868.2655045} contains 8732 labeled urban sound recordings in .wav format. All recordings are of a duration of 4 seconds from 10 classes. The files are sorted by 10 folds (folders called fold1-fold10)

\subsection{Pre-processing}
First, we apply augmentation on signal level, as the mentioned augmentation methods perform better on signal level rather than mel-spectrogram. We resize the mel-spectrogram to 32 x 32 as an image-like feature before feeding to the network.  RandAug applied before training as a data preprocessing step.  



\section{Results}

To evaluate the effectiveness of our proposed approach, we conducted experiments using various models on two datasets: FSDD and UrbanSound8K. It is important to note we included all the techniques in the search space that perform better than the baseline.  We present the experimental results in Table~\ref{tab:custom_cnn}, where a custom CNN was used as the baseline for both datasets. Accuracy served as the evaluation metric. The table reports the difference between each data augmentation (DA) technique and the baseline accuracy, denoted as $\Delta D$. A green $\Delta D$ indicates an improvement in accuracy compared to the baseline, while a red $\Delta D$ signifies a decrease. Only the data augmentation techniques that demonstrated improved accuracy are included in the table.

Our proposed data augmentation technique using custom CNN exhibited a significant absolute improvement of 5.16\% on the FSDD dataset and 1.37\% on the UrbanSound8K dataset. The 5.16\% absolute improvement over the baseline on the FSDD dataset is particularly noteworthy, as it represents the highest accuracy performance among all the utilized DAs. Although the 1.37\% improvement on the UrbanSound8K dataset is not the highest, it still demonstrates a competitive enhancement in accuracy.

For the pre-trained VGG model, we conducted a similar set of experiments as with the CNN model. However, we observed that fewer data augmentation methods showed improved performance compared to the CNN case. Therefore, we excluded those augmentations with lower accuracy performance compared to the baseline from the search space. Our proposed method exhibited superior accuracy performance compared to all other data augmentation methods across both datasets. For the FSDD dataset, the proposed method showed an absolute improvement of nearly 3\% over the baseline, while for the UrbanSound8K dataset, it demonstrated an absolute improvement of approximately 2.30\%. Overall, our proposed method achieved the best accuracy performance among all the methods employed.


\section{Conclusion}\label{conclusion}
This paper introduces AudRandAug, a novel data augmentation technique specifically designed for audio data. AudRandAug selects data augmentation policies from a dedicated audio search space and demonstrates remarkable performance improvements compared to the baseline. Through extensive experiments on FSDD and UrbanSound8K datasets, using various models, AudRandAug consistently outperforms other data augmentation methods.    The results validate the effectiveness and potential of AudRandAug in enhancing the performance of audio-related models. By addressing the specific needs of audio data, this research contributes to the advancement of audio tasks within the computer vision field. 
In future, AudRandAug can be used as a powerful technique for audio data augmentation, demonstrating significant accuracy improvements. This work opens up possibilities for further research and development of tailored data augmentation methods to optimize audio-related applications.
\section{Acknowledgment}
This research was supported by Science Foundation Ireland under grant numbers 18/CRT/6223 (SFI Centre for Research Training in Artificial intelligence) and  13/RC/2094/P\_2 (Lero SFI Centre for Software). For the purpose of Open Access, the author has applied a CC BY public copyright licence to any Author Accepted Manuscript version arising from this submission.


\begin{thebibliography}{00}


\bibitem{dong2018convolutional}Dong, M. Convolutional neural network achieves human-level accuracy in music genre classification. {\em ArXiv Preprint ArXiv:1802.09697}. (2018)
\bibitem{choi2017convolutional}Choi, K., Fazekas, G., Sandler, M. \& Cho, K. Convolutional recurrent neural networks for music classification. {\em 2017 IEEE International Conference On Acoustics, Speech And Signal Processing (ICASSP)}. pp. 2392-2396 (2017)
\bibitem{zhang2016improved}Zhang, W., Lei, W., Xu, X. \& Xing, X. Improved music genre classification with convolutional neural networks.. {\em Interspeech}. pp. 3304-3308 (2016)
\bibitem{aleem2022random}Aleem, S., Kumar, T., Little, S., Bendechache, M., Brennan, R. \& McGuinness, K. Random data augmentation based enhancement: a generalized enhancement approach for medical datasets. {\em ArXiv Preprint ArXiv:2210.00824}. (2022)
\bibitem{oord2016wavenet}Oord, A., Dieleman, S., Zen, H., Simonyan, K., Vinyals, O., Graves, A., Kalchbrenner, N., Senior, A. \& Kavukcuoglu, K. Wavenet: A generative model for raw audio. {\em ArXiv Preprint ArXiv:1609.03499}. (2016)
\bibitem{roberts2018hierarchical}Roberts, A., Engel, J., Raffel, C., Hawthorne, C. \& Eck, D. A hierarchical latent vector model for learning long-term structure in music. {\em International Conference On Machine Learning}. pp. 4364-4373 (2018)
\bibitem{guzhov2021esresnet}Guzhov, A., Raue, F., Hees, J. \& Dengel, A. Esresnet: Environmental sound classification based on visual domain models. {\em 2020 25th International Conference On Pattern Recognition (ICPR)}. pp. 4933-4940 (2021)
\bibitem{aytar2016soundnet}Aytar, Y., Vondrick, C. \& Torralba, A. Soundnet: Learning sound representations from unlabeled video. {\em Advances In Neural Information Processing Systems}. \textbf{29} (2016)
\bibitem{demir2020new}Demir, F., Abdullah, D. \& Sengur, A. A new deep CNN model for environmental sound classification. {\em IEEE Access}. \textbf{8} pp. 66529-66537 (2020)
\bibitem{tokozume2017learning}Tokozume, Y. \& Harada, T. Learning environmental sounds with end-to-end convolutional neural network. {\em 2017 IEEE International Conference On Acoustics, Speech And Signal Processing (ICASSP)}. pp. 2721-2725 (2017)
\bibitem{lee2017sample}Lee, J., Park, J., Kim, K. \& Nam, J. Sample-level deep convolutional neural networks for music auto-tagging using raw waveforms. {\em ArXiv Preprint ArXiv:1703.01789}. (2017)
\bibitem{chachada2014environmental}Chachada, S. \& Kuo, C. Environmental sound recognition: A survey. {\em APSIPA Transactions On Signal And Information Processing}. \textbf{3} (2014)
\bibitem{dandashi2017survey}Dandashi, A. \& AlJaam, J. A survey on audio content-based classification. {\em 2017 International Conference On Computational Science And Computational Intelligence (CSCI)}. pp. 408-413 (2017)
\bibitem{palanisamy2020rethinking}Palanisamy, K., Singhania, D. \& Yao, A. Rethinking CNN models for audio classification. {\em ArXiv Preprint ArXiv:2007.11154}. (2020)
\bibitem{jackson2016free}Jackson, Z. Free spoken digit dataset (fsdd). {\em Retrieved February}. \textbf{1} pp. 2020 (2016)
\bibitem{BSGHC3_2020_v25n6_854}Park, J., Kumar, T. \& Bae, S. Search for Optimal Data Augmentation Policy for Environmental Sound Classification with Deep Neural Networks. {\em Journal Of Broadcast Engineering}. \textbf{6} (2020,11), http://dx.doi.org/10.5909/JBE.2020.25.6.854
\bibitem{kumar2020intra}Kumar, T., Park, J. \& Bae, S. Intra-Class Random Erasing (ICRE) augmentation for audio classification. {\em Proceedings Of The Korean Society Of Broadcast Engineers Conference}. pp. 244-247 (2020)
\bibitem{salamon2017deep}Salamon, J. \& Bello, J. Deep convolutional neural networks and data augmentation for environmental sound classification. {\em IEEE Signal Processing Letters}. \textbf{24}, 279-283 (2017)
\bibitem{wu2015image}Wu, M. \& Chen, L. Image recognition based on deep learning. {\em 2015 Chinese Automation Congress (CAC)}. pp. 542-546 (2015)
\bibitem{fujiyoshi2019deep}Fujiyoshi, H., Hirakawa, T. \& Yamashita, T. Deep learning-based image recognition for autonomous driving. {\em IATSS Research}. \textbf{43}, 244-252 (2019)
\bibitem{ranjbarzadeh2023me}Ranjbarzadeh, R., Jafarzadeh Ghoushchi, S., Tataei Sarshar, N., Tirkolaee, E., Ali, S., Kumar, T. \& Bendechache, M. ME-CCNN: Multi-encoded images and a cascade convolutional neural network for breast tumor segmentation and recognition. {\em Artificial Intelligence Review}. pp. 1-38 (2023)
\bibitem{kumar2023rsmda}Kumar, T., Mileo, A., Brennan, R. \& Bendechache, M. RSMDA: Random Slices Mixing Data Augmentation. {\em Applied Sciences}. \textbf{13}, 1711 (2023)
\bibitem{kumar2023advanced}Kumar, T., Turab, M., Raj, K., Mileo, A., Brennan, R. \& Bendechache, M. Advanced Data Augmentation Approaches: A Comprehensive Survey and Future directions. {\em ArXiv Preprint ArXiv:2301.02830}. (2023)
\bibitem{kumar2021class}Kumar, T., Park, J., Ali, M., Uddin, A. \& Bae, S. Class Specific Autoencoders Enhance Sample Diversity. {\em Journal Of Broadcast Engineering}. \textbf{26}, 844-854 (2021)
\bibitem{kumar2021binary}Kumar, T., Park, J., Ali, M., Uddin, A., Ko, J. \& Bae, S. Binary-classifiers-enabled filters for semi-supervised learning. {\em IEEE Access}. \textbf{9} pp. 167663-167673 (2021)
\bibitem{torfi2020natural}Torfi, A., Shirvani, R., Keneshloo, Y., Tavaf, N. \& Fox, E. Natural language processing advancements by deep learning: A survey. {\em ArXiv Preprint ArXiv:2003.01200}. (2020)
\bibitem{hirschberg2015advances}Hirschberg, J. \& Manning, C. Advances in natural language processing. {\em Science}. \textbf{349}, 261-266 (2015)
\bibitem{hershey2017cnn}Hershey, S., Chaudhuri, S., Ellis, D., Gemmeke, J., Jansen, A., Moore, R., Plakal, M., Platt, D., Saurous, R., Seybold, B. \& Others CNN architectures for large-scale audio classification. {\em 2017 Ieee International Conference On Acoustics, Speech And Signal Processing (icassp)}. pp. 131-135 (2017)
\bibitem{fu2010survey}Fu, Z., Lu, G., Ting, K. \& Zhang, D. A survey of audio-based music classification and annotation. {\em IEEE Transactions On Multimedia}. \textbf{13}, 303-319 (2010)
\bibitem{chandio2021audd}Chandio, A., Shen, Y., Bendechache, M., Inayat, I. \& Kumar, T. AUDD: audio Urdu digits dataset for automatic audio Urdu digit recognition. {\em Applied Sciences}. \textbf{11}, 8842 (2021)
\bibitem{park2020search}Park, J., Kumar, T. \& Bae, S. Search for optimal data augmentation policy for environmental sound classification with deep neural networks. {\em Journal Of Broadcast Engineering}. \textbf{25}, 854-860 (2020)
\bibitem{turab2022investigating}Turab, M., Kumar, T., Bendechache, M. \& Saber, T. Investigating Multi-Feature Selection and Ensembling for Audio Classification. {\em ArXiv Preprint ArXiv:2206.07511}. (2022)
\bibitem{cubuk2020randaugment}Cubuk, E., Zoph, B., Shlens, J. \& Le, Q. Randaugment: Practical automated data augmentation with a reduced search space. {\em Proceedings Of The IEEE/CVF Conference On Computer Vision And Pattern Recognition Workshops}. pp. 702-703 (2020)
\bibitem{li2019multi}Li, X., Chebiyyam, V. \& Kirchhoff, K. Multi-stream network with temporal attention for environmental sound classification. {\em ArXiv Preprint ArXiv:1901.08608}. (2019)
\bibitem{ko2015audio}Ko, T., Peddinti, V., Povey, D. \& Khudanpur, S. Audio augmentation for speech recognition. {\em Sixteenth Annual Conference Of The International Speech Communication Association}. (2015)
\bibitem{jain2021spliceout}Jain, A., Samala, P., Mittal, D., Jyoti, P. \& Singh, M. Spliceout: A simple and efficient audio augmentation method. {\em ArXiv Preprint ArXiv:2110.00046}. (2021)
\bibitem{lee2017raw}Lee, J., Kim, T., Park, J. \& Nam, J. Raw waveform-based audio classification using sample-level CNN architectures. {\em ArXiv Preprint ArXiv:1712.00866}. (2017)
\bibitem{nanni2020data}Nanni, L., Maguolo, G. \& Paci, M. Data augmentation approaches for improving animal audio classification. {\em Ecological Informatics}. \textbf{57} pp. 101084 (2020)
\bibitem{su2019environment}Su, Y., Zhang, K., Wang, J. \& Madani, K. Environment sound classification using a two-stream CNN based on decision-level fusion. {\em Sensors}. \textbf{19}, 1733 (2019)
\bibitem{liu1998audio}Liu, Z., Wang, Y. \& Chen, T. Audio feature extraction and analysis for scene segmentation and classification. {\em Journal Of VLSI Signal Processing Systems For Signal, Image And Video Technology}. \textbf{20}, 61-79 (1998)
\bibitem{kim2021specmix}Kim, G., Han, D. \& Ko, H. SpecMix: A mixed sample data augmentation method for training withtime-frequency domain features. {\em ArXiv Preprint ArXiv:2108.03020}. (2021)
\bibitem{10.1145/2647868.2655045}Salamon, J., Jacoby, C. \& Bello, J. A Dataset and Taxonomy for Urban Sound Research. {\em Proceedings Of The 22nd ACM International Conference On Multimedia}. pp. 1041-1044 (2014), https://doi.org/10.1145/2647868.2655045
\bibitem{roy2023wildect}Roy, A., Bhaduri, J., Kumar, T. \& Raj, K. WilDect-YOLO: An efficient and robust computer vision-based accurate object localization model for automated endangered wildlife detection. {\em Ecological Informatics}. \textbf{75} pp. 101919 (2023)
\bibitem{khan2022introducing}Khan, W., Raj, K., Kumar, T., Roy, A. \& Luo, B. Introducing urdu digits dataset with demonstration of an efficient and robust noisy decoder-based pseudo example generator. {\em Symmetry}. \textbf{14}, 1976 (2022)
\bibitem{chandio2022precise}Chandio, A., Gui, G., Kumar, T., Ullah, I., Ranjbarzadeh, R., Roy, A., Hussain, A. \& Shen, Y. Precise single-stage detector. {\em ArXiv Preprint ArXiv:2210.04252}. (2022)
\bibitem{singh2023deep}Singh, A., Raj, K., Kumar, T., Verma, S. \& Roy, A. Deep learning-based cost-effective and responsive robot for autism treatment. {\em Drones}. \textbf{7}, 81 (2023)
\bibitem{singh2023understanding}Singh, A., Ranjbarzadeh, R., Raj, K., Kumar, T. \& Roy, A. Understanding EEG signals for subject-wise definition of armoni activities. {\em ArXiv Preprint ArXiv:2301.00948}. (2023)
\bibitem{roy2022computer}Roy, A., Bhaduri, J., Kumar, T. \& Raj, K. A computer vision-based object localization model for endangered wildlife detection. {\em Ecological Economics, Forthcoming}. (2022)
\bibitem{kumar2022forged}Kumar, T., Turab, M., Talpur, S., Brennan, R. \& Bendechache, M. Forged character detection datasets: passports, driving licences and visa stickers. {\em Int. J. Artif. Intell. Appl.(IJAIA)}. \textbf{13} pp. 21-35 (2022)
\bibitem{kumar2022stride}Kumar, T., Brennan, R. \& Bendechache, M. Stride Random Erasing Augmentation. {\em CS \& IT Conference Proceedings}. (2022)
\bibitem{turab2023comprehensive}Turab, M. \& Jamil, S. A Comprehensive Survey of Digital Twins in Healthcare in the Era of Metaverse. {\em BioMedInformatics}. \textbf{3}, 563-584 (2023)
\bibitem{sarwar2022advanced}Sarwar, S., Turab, M., Channa, D., Chandio, A., Sohu, M. \& Kumar, V. Advanced Audio Aid for Blind People. {\em 2022 International Conference On Emerging Technologies In Electronics, Computing And Communication (ICETECC)}. pp. 1-6 (2022)
\bibitem{khan2022data}Khan, W., Turab, M., Ahmad, W., Ahmad, S., Kumar, K. \& Luo, B. Data Dimension Reduction makes ML Algorithms efficient. {\em 2022 International Conference On Emerging Technologies In Electronics, Computing And Communication (ICETECC)}. pp. 1-7 (2022)
\bibitem{irfan2023go}Irfan, A., Nawaz, A., Turab, M., Azeem, M., Adnan, M., Mehmood, A., Ahmed, S. \& Ashraf, A. Go Together: Bridging the Gap between Learners and Teachers. {\em 2023 7th International Multi-Topic ICT Conference (IMTIC)}. pp. 1-7 (2023)
\bibitem{steinberg2010method}Steinberg, E., Corcoran, P. \& Prilutsky, Y. Method of determining PSF using multiple instances of nominally scene. (Google Patents,2010), US Patent 7,660,478
\bibitem{mandal2023measuring}Mandal, A., Leavy, S. \& Little, S. Measuring Bias in Multimodal Models: Multimodal Composite Association Score. {\em International Workshop On Algorithmic Bias In Search And Recommendation}. pp. 17-30 (2023)
\bibitem{mandal2023multimodal}Mandal, A., Leavy, S. \& Little, S. Multimodal Composite Association Score: Measuring Gender Bias in Generative Multimodal Models. {\em ArXiv Preprint ArXiv:2304.13855}. (2023)
\bibitem{mandal2021dataset}Mandal, A., Leavy, S. \& Little, S. Dataset diversity: measuring and mitigating geographical bias in image search and retrieval. {\em Proceedings Of The 1st International Workshop On Trustworthy AI For Multimedia Computing}. pp. 19-25 (2021)
\bibitem{mandal2016static}Mandal, A., Sharma, U. \& Pant, H. Static structural analysis of universal joint to study the various stresses and strains developed in power transmission systems. {\em International Journal Of Engineering Research \& Technology}. \textbf{5} (2016)
\end{thebibliography}
\end{document}